\documentclass{ws-procs975x65}  
\usepackage{graphicx}% Include figure files
\usepackage{subfigure}
\newcommand{\prd}{Phys. Rev. D}

\begin{document}
\title{Bound orbit domains in the phase space of the Kerr geometry}

\author{Prerna Rana$^{1, \dagger}$ and  A.\ Mangalam$^{2, \ddagger}$}

\address{$^1$Department of Astronomy and Astrophysics, Tata Institute of Fundamental Research, Homi Bhabha Road, Navy Nagar, Colaba, Mumbai 400005, India \\ $^2$ Indian Institute of Astrophysics, Sarjapur Road, 2nd Block Koramangala, Bangalore, 560034, India \\
E-mail: prerna.rana@tifr.res.in$^\dagger$, mangalam@iiap.res.in$^\ddagger$}

\begin{abstract}
We derive the conditions for a non-equatorial eccentric bound orbit to exist around a Kerr black hole in two-parameter spaces: the energy, angular momentum of the test particle, spin of the black hole, and Carter's constant space ($E$, $L$, $a$, $Q$), and eccentricity, inverse-latus rectum space ($e$, $\mu$, $a$, $Q$). These conditions distribute various kinds of bound orbits in different regions of the ($E$, $L$) and ($e$, $\mu$) planes, depending on which pair of roots of the effective potential forms a bound orbit. We provide a prescription to select these parameters for bound orbits, which are useful inputs to study bound trajectory evolution in various astrophysical applications like simulations of gravitational wave emission from extreme-mass ratio inspirals, relativistic precession around black holes, and the study of gyroscope precession as a test of general relativity. 
\end{abstract}


\bodymatter

%%%%%%%%%%%%%%%%% now a standard article style for the most part

\section{Introduction}
Bound trajectories in the Kerr geometry have been studied extensively, and some of the important results are discussed in a pioneering work by S. Chandrasekhar (Ref.~\refcite{C1983book}). The general trajectory in the Kerr spacetime was first expressed in terms of quadratures in Ref.~\refcite{Carter1968}, while Ref.~\refcite{Wilkins1972} discusses the essential conditions for bound spherical geodesics, and also horizon-skimming orbits. The quadrature form of the fundamental orbital frequencies for a general eccentric trajectory was first presented in Ref.~\refcite{Schmidt2002}. Later, to decouple the $r$ and $\theta$ motions, a parameter called Mino time, $\lambda$, was introduced in Ref.~\refcite{Mino2003}, which was then implemented to calculate a closed-form trajectory solution and orbital frequencies in Ref.~\refcite{Fujita2009}. Recently, an alternate analytic solution was derived for the general bound and separatrix trajectories in a compact form using the transformation $1/r = \mu \left(1+ e \cos \chi \right)$ in Refs.~(\refcite{RMCQG2019,RM1arxiv2019}). The inputs to these integrals for calculating the trajectories are the constants of motion $E$, $L$, $Q$, and spin of the black hole, $a$. These parameters can also be translated to ($e$, $\mu$, $a$, $Q$) space, as derived in Refs.~(\refcite{RMCQG2019,RM1arxiv2019}). It is essential to find the canonical bound orbit conditions in these two parameter spaces to calculate the trajectory evolution. 

We express the bound orbit conditions on ($E$, $L$, $a$, $Q$) parameters for the non-equatorial eccentric bound orbits around a Kerr black hole, and then find the analog of these conditions in the ($e$, $\mu$, $a$, $Q$) space. The regions of different bound orbits were graphically separated in the ($E$, $L$) plane in Ref.~\refcite{Hackmann2010}, according to the pair of roots of the effective potential spanning the radius of the bound orbit. It is essential to find the canonical bound orbit conditions in these two parameter spaces.

\section{Conditions for bound trajectories around Kerr black hole}
 Now, we consider the radial motion of the bound trajectory which is described by the equation (Refs. \refcite{Carter1968,Schmidt2002,RMCQG2019,RM1arxiv2019})
\begin{equation}
\frac{\left( E^2 - 1 \right)}{2}=\frac{ \rho^{4}}{2 r^4}\left( \frac{{\rm d}r}{{\rm d} \tau}\right)^2 -\frac{1}{r}+ \frac{L^2-a^2\left( E^2 -1\right)+Q }{2 r^2}-\frac{\left( L- aE\right)^2 +Q }{r^3}+\frac{a^2 Q}{2r^4}, \label{radialeqn1}
\end{equation}
to derive the conditions in ($E, L, a, Q$) and ($e, \mu, a, Q$) spaces for various bound orbit regions previously discussed in Ref.~\refcite{Hackmann2010}, where $\rho^{2}=r^2 +a^2 \cos^2 \theta$, $\tau$ is the proper time, and $a=J/M^2$ which have their usual meanings; we use geometrical units throughout. 

\subsection{\underline{Dynamical parameter space ($E, L, a, Q$)}}
\label{ELaQ}
Next, we solve the quartic equation, Eq. (\ref{radialeqn1}), to find the four roots (that include the turning points of the bound orbit) of ${\rm d}r / {\rm d} \tau=0$, which can be expressed as
 \begin{eqnarray}
r^4 + \dfrac{2}{\left( E^2 -1\right)}r^3 + \dfrac{\left( a^2 E^2 -L^2 -Q -a^2\right)}{\left( E^2 -1\right)}r^2 + \dfrac{2 \left(x^{2}+Q \right)}{\left( E^2 -1\right)}r -\dfrac{a^2 Q}{\left( E^2 -1\right)}=0, \label{quarticr}
\end{eqnarray}
where $x=L-aE$. Applying Ferrari's method (Ref.~\refcite{Ferrari}) to the above equation gives
 \begin{subequations}
 \begin{eqnarray}
 r_{1}=&&\dfrac{1}{2\left(1- E^2 \right)}+\dfrac{\sqrt{2z}}{2}+\dfrac{1}{2}\sqrt{D_1}, \label{rt1}\\
  r_{2}=&&\dfrac{1}{2\left( 1-E^2 \right)}+\dfrac{\sqrt{2z}}{2}-\dfrac{1}{2}\sqrt{D_1}, \label{rt2}\\
   r_{3}=&&\dfrac{1}{2\left(1- E^2 \right)}-\dfrac{\sqrt{2z}}{2}+\dfrac{1}{2}\sqrt{D_2}, \label{rt3} \\
     r_{4}=&&\dfrac{1}{2\left(1- E^2\right)}-\dfrac{\sqrt{2z}}{2}-\dfrac{1}{2}\sqrt{D_2}, \label{rt4}
 \end{eqnarray}
 where $r_1 > r_2 > r_3 > r_4$, and
 \begin{eqnarray}
  D_1=&&-2G-2z-\dfrac{\sqrt{2}H}{\sqrt{z}}, \ \  D_2=-2G-2z+\dfrac{\sqrt{2}H}{\sqrt{z}}, \\
   z=&&U+V-\dfrac{G}{3}, \ \ \
 U=\left( I+\sqrt{I^2+J^3} \right) ^{1/3}, \ \ \
  V=\left( I-\sqrt{I^2+J^3} \right) ^{1/3}, \\
  I=&&\dfrac{\left( 2G^3 +27 H^2 -72 G T\right) }{432}, \ \
  J= -\dfrac{\left( G^2 +12 T\right) }{36}, \label{J} \\
   G=&&\dfrac{\left[ L^2 -a^2 \left( E^2 -1\right)+Q \right] }{\left( 1-E^2\right) }-\dfrac{3}{2\left(1- E^2 \right)^2}, \\
     H=&&\dfrac{\left[ L^2-a^2\left( E^2 -1\right)+Q\right] }{\left(1- E^2 \right)^2}-\dfrac{2\left( x^2 +Q \right) }{\left(1- E^2 \right)}-\dfrac{1}{\left(1- E^2 \right)^3}, \\
 T=&&\dfrac{\left[L^2 -a^2 \left( E^2 -1\right)+Q \right] }{4\left(1- E^2 \right)^3}-\dfrac{3}{16\left(1- E^2 \right)^4}-\dfrac{\left( x^2+ Q\right) }{\left( 1-E^2\right)^2}+\dfrac{a^2 Q}{\left( 1-E^2 \right)}.
 \end{eqnarray}
 \label{rootseqns}
 \end{subequations}
 The bound orbit regions were graphically classified in the ($E, L$) plane in Ref.~\refcite{Hackmann2010} on the basis of which pair of roots of quartic equation, Eq. (\ref{quarticr}), contains the bound orbit. We algebraically classify these regions using the expressions of roots, Eqs. (\ref{rt1}-\ref{rt4}), in the ($E, L, a, Q$) parameter space as follows (Ref.~\refcite{RM2019}):
 \begin{enumerate}
 \item \underline{\textbf{Region $\Delta$}}: Bound orbits exist between ($r_1$, $r_2$), and between ($r_3$, $r_4$): $D_1>0$, $D_2>0$, and $E<1$. 
 \item \underline{\textbf{Region $\varsigma$}}: Bound orbit either exists between $r_1$ and $r_2$ if ($r_3$, $r_4$) forms a complex pair i.e. $D_2<0$ or exists between $r_3$ and $r_4$ if ($r_1$, $r_2$) forms a complex pair i.e. $D_1<0$: $\left( D_1 \cdot D_2\right)  < 0$. This region exists for both $E < 1$ and $E > 1$.
 \item \underline{\textbf{Region $\Lambda$}}:  Bound orbit exists between $r_2$ and $r_3$ with $r_4<r_3$ and $r_1 > r_2$: $D_1>0$, $D_2>0$, and $E>1$.
 \end{enumerate}
  \begin{figure}[ht!]
\begin{center}
\hspace{0.5cm}
\includegraphics[scale=0.45]{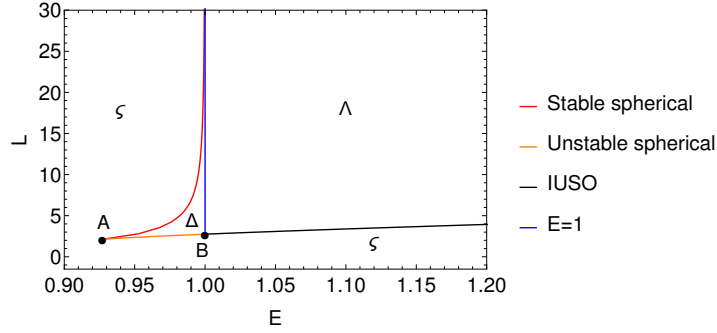}
\end{center}
\caption{\label{ELplane}The bound orbit regions $\Delta$, $\varsigma$, and $\Lambda$ are shown in the ($E$, $L$) plane for $a=0.5$, $Q=5$, where the region $\Delta$ is bounded by stable and  unstable (also corresponds to the separatrices) spherical orbits, and $E=1$; the $\Lambda$ region is bounded by the inner unstable spherical orbits (IUSO). Points A and B represent the innermost stable spherical orbit (ISSO) and marginally stable spherical orbit (MBSO) respectively.}
\end{figure}
The classification of these regions in the ($E$, $L$) plane is shown in Fig. \ref{ELplane}. The bounding curves of these regions represent spherical orbits. The eccentricity and inverse latus-rectum of the bound orbit are defined as (Ref.~\refcite{RM2019})
 \begin{eqnarray}
 e_{ij}=\dfrac{r_i -r_j}{r_i + r_j}, \ \
 \mu_{ij}=\dfrac{r_i + r_j}{2 r_i r_j}, \label{genmu}
 \end{eqnarray}
 where we see that \{$e_{ij}$, $\mu_{ij}$\} can be expressed in terms of ($E$, $L$, $a$, $Q$) through roots, Eqs. \eqref{rootseqns}. The details of derivations presented here are provided in Ref.~\refcite{RM2019}.

\subsection{\underline{Conic parameter space ($e, \mu, a, Q$)}}
\label{emuaQ}
According to the definitions given for regions $\Delta$, $\varsigma$, and $\Lambda$ in \S \ref{ELaQ}, the convention adopted for \{$e$, $\mu$\} is as follows: Region $\Delta$ given by \{$e_{12}$, $\mu_{12}$\}; Region $\Lambda$ given by \{$e_{23}$, $\mu_{23}$\}; Region $\varsigma$ given by \{$e_{12}$, $\mu_{12}$\} or \{$e_{34}$, $\mu_{34}$\}, depending on which pair is real.
Now, we derive the defining conditions for $\Delta$, $\varsigma$, and $\Lambda$ regions  in the ($e$, $\mu$, $a$, $Q$) space:

(i) \underline{\textbf{Region $\Delta$}}: The defining conditions for this region was derived using the necessary constraints on the parameters of the Elliptic integrals involved in the trajectory solutions (Refs.~\refcite{RMCQG2019,RM1arxiv2019}), which are
\begin{subequations}
\begin{eqnarray}
\left[ \mu^3 a^2 Q \left(1+e \right)^2 +\mu^2 \left( \mu a^2 Q -x^2 -Q\right) \left(3-e \right) \left(1+e \right) +1 \right]  > 0, \label{deltaregiona}\\
\left[ \mu \left( 1+e\right) \left( 1+ \sqrt{1-a^2}\right) \right]  <1, \label{deltaregionb} \\
E(e, \mu, a, Q) < 1. \label{deltaregionc}
\end{eqnarray}
\label{deltaregion}
\end{subequations}
(ii) \underline{\textbf{Region $\varsigma$}}: The region $\varsigma$ is defined by two complex roots of $r$ or $u=1/r$ with a bound orbit existing between the two remaining real roots. We can write Eq. (\ref{quarticr}) for region $\varsigma$ in the form (Ref.~\refcite{RM2019})
\begin{subequations}
\begin{equation}
\left[u - \mu \left(1-e \right)  \right] \cdot \left[u - \mu \left(1+e \right)  \right] \cdot \left(u^2 + Au +B \right)=0, \label{uform}
\end{equation}
where bound orbit exists between $\mu \left(1+e \right)$ and $\mu \left(1-e \right)$ which is a real pair of the roots, and
\begin{equation}
A=- 2\left[ \dfrac{x^2 +Q}{a^2 Q}- \mu \right], \ \ \  B=\dfrac{1-E^2}{\mu^2 a^2 Q \left( 1-e^2\right) }.
\end{equation} 
\end{subequations}
Hence, the remaining factor $\left(u^2 + Au +B \right)$ of Eq. (\ref{uform}) should have complex roots for the $\varsigma$ region, which reduces to the condition (Ref.~\refcite{RM2019})
\begin{equation}
 \left\lbrace  \left( Q + x^2 \right)^2 \mu + a^4 e^2 Q^2 \mu^3 - a^2 Q \left[1 + \left( 1+ e^2\right) \left( Q +x^2\right) \mu^2   \right] \right\rbrace <0, \label{boundcondreg5}
\end{equation}
where we have substituted for the factor ($1-E^2$) in terms of ($e$, $\mu$, $x$) using relations previously derived in Refs.~(\refcite{RMCQG2019,RM1arxiv2019}).

\begin{figure}[ht!]
\begin{center}
\mbox{ \subfigure[]{
\includegraphics[width=6.3cm,height=3.8cm]{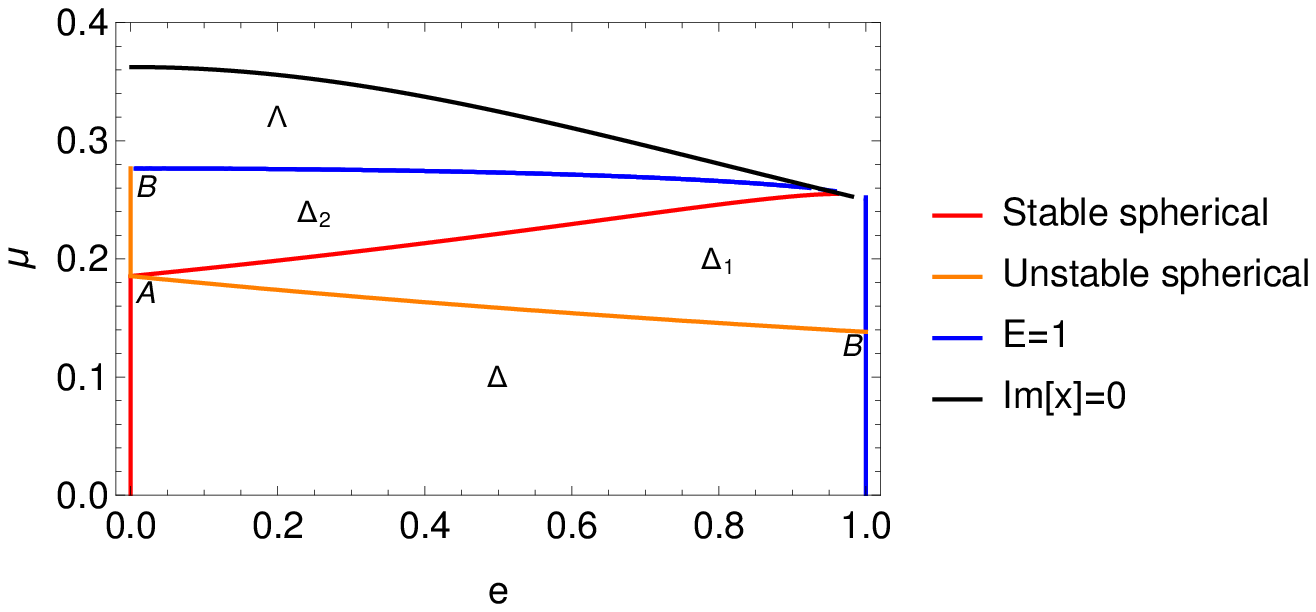} \label{emuQ2a2reg5}}
\qquad
\subfigure[]{
\includegraphics[width=6.3cm,height=3.8cm]{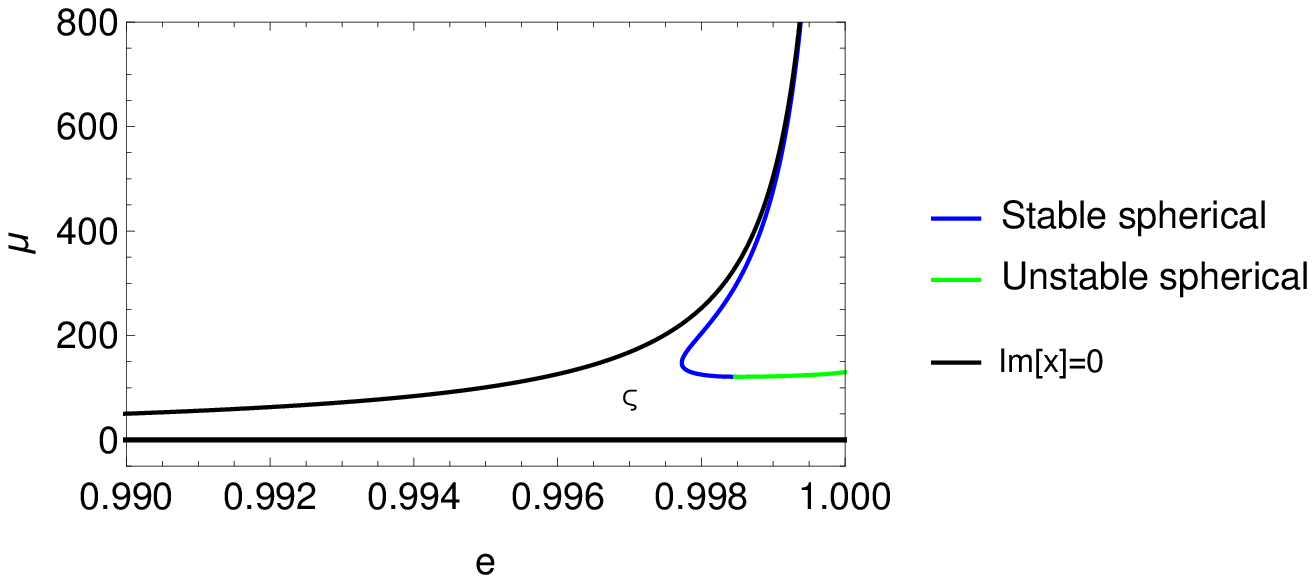} \label{emuQ2a2reg4}}}
\end{center}
\caption{\label{emuQ2a2}The bound orbit regions (a) $\Delta$ and $\Lambda$, and (b) $\varsigma$ in the ($e$, $\mu$) plane for ($a=0.2$, $Q=2$) are shown along with their defining bounding curves and end points. The regions $\Delta_1$ and $\Delta_2$ are replicas of region $\Delta$ when ($e_{13}$, $\mu_{13}$) and ($e_{23}$, $\mu_{23}$) are chosen respectively.}
\end{figure}

(iii) \underline{\textbf{Region $\Lambda$}}: The region $\Lambda$ is defined by the condition that a bound orbit exists between $r_2$ and $r_3$ (or $u_2$ and $u_3$) with $r_1 > r_2$ (or $u_1 < u_2$) and $r_3 > r_4$ (or $u_3 < u_4$). We can express Eq. (\ref{quarticr}) for this region as
\begin{equation}
\left[u - \mu \left(1-e \right)  \right] \cdot \left[u - \mu \left(1+e \right)  \right] \cdot \left(u^2 + Au +B \right)=0. \label{uformlambda}
\end{equation}
The remaining roots $u_1$ and $u_4$ can be derived from the factor $\left(u^2 + Au +B \right)$, which can be substituted into the conditions $u_1 < u_2$ and $u_3 < u_4$ to obtain
\begin{subequations}
\begin{equation}
\left[ \mu^3 a^2 Q \left(1-e \right)^2 +\mu^2 \left( \mu a^2 Q -x^2 -Q\right) \left(3+e \right) \left(1-e \right) +1 \right]  < 0,  \label{boundcondreg41}
\end{equation}
and
\begin{equation}
\left[ \mu^3 a^2 Q \left(1+e \right)^2 +\mu^2 \left( \mu a^2 Q -x^2 -Q\right) \left(3-e \right) \left(1+e \right) +1 \right]  < 0,  \label{boundcondreg42}
\end{equation}
 respectively. Next, as we see from Fig. \ref{ELplane}, that the region $\Lambda$ corresponds to the orbits with $E>1$; this implies
\begin{equation}
\left[ \mu^2 \left( 1-e^2\right) \left( \mu a^2 Q -Q -x^2\right) +1\right] <0. \label{boundcondreg4}
\end{equation}
 \end{subequations}
 In effect, Eqs. (\ref{boundcondreg41}, \ref{boundcondreg42}, \ref{boundcondreg4}) together define $\Lambda$ region the ($e$, $\mu$) plane, and Fig. \ref{emuQ2a2} shows all these regions in the ($e$, $\mu$) plane. The details of derivations of the above conditions are provided in Ref.~\refcite{RM2019}.

 \section{A prescription for selecting bound orbits}
\label{xybound}
We present the scaling formulae specifying the parameters ($E$, $L$) and ($e$, $\mu$) in terms of the variables ($x_{1}$, $y_{1}$) and ($x_{2}$, $y_{2}$), where $x_i$, $y_i\in \left[ 0, 1 \right]$, $i=1, 2$, can be chosen to produce valid combinations of the parameters ($E$, $L$) and ($e$, $\mu$) for bound orbits.
The formula for selecting $E$ for bound orbits written in terms of the variable $x_{1}$, $a$, and $Q$ for region $\Delta$ in Fig. \ref{ELplane} is (Ref.~\refcite{RM2019})
\begin{equation}
E\left(x_{1}, a ,Q\right) =E_{Z}\left(a, Q \right)  + x_{1}\left[E_{Y}\left(a, Q \right)- E_{Z}\left(a, Q \right)\right], \label{Exa}
\end{equation} 
where $E_{Z}\left(a, Q \right)$ and $E_{Y}\left(a, Q \right)$ are the spherical orbit energies at ISSO and MBSO respectively, and where $Z(a, Q)$ and $Y(a, Q)$ are radii of ISSO and MBSO respectively (Eqs. (19), (20) in Refs.~\refcite{RMCQG2019,RM1arxiv2019}).

Now, for a fixed $x_{1}$ and $a$, $y_{1} \in \left[ 0, 1 \right]$ defines the range of $L\left( r,a, Q\right)$ for bound orbits. The formula for selecting allowed $L$ ($\Delta$ in Fig. \ref{ELplane}) can be written as (Ref.~\refcite{RM2019})
\begin{subequations}
\begin{equation}
\dfrac{1}{L \left( x_{1}, y_{1}, a, Q \right)}= \dfrac{1}{L_{-}\left( x_{1}, a, Q\right) } - y_{1}\left[\dfrac{1}{L_{-}\left( x_{1}, a, Q\right)}-\dfrac{1}{L_{+}\left( x_{1}, a, Q\right)} \right], \label{Lxya}
\end{equation}
where $L_{+}\left( x_{1}, a, Q\right)$ and $L_{-}\left( x_{1}, a, Q\right)$ are end points of the $\Delta$ region in Fig. \ref{ELplane} given by (Ref.~\refcite{RM2019})
\begin{eqnarray}
L_{+}\left( x_{1}, a, Q\right)&&=x(r_{v}\left(x_{1}, a, Q \right),a,Q)+a \cdot E(r_{v}\left(x_{1}, a, Q \right),a,Q), \label{L+} \\
L_{-}\left( x_{1}, a, Q\right)&&=x(r_{u}\left(x_{1}, a, Q \right),a,Q)+a \cdot E(r_{u}\left(x_{1}, a, Q \right),a,Q), \label{L-}
\end{eqnarray}
\end{subequations}
where $x\left( r_s ,a, Q\right)$ and $E\left( r_s ,a, Q\right)$ can be calculated using the spherical orbit formulae (derived in Refs.~\refcite{RMCQG2019,RM1arxiv2019}) and where $r_{v}\left(x_{1}, a , Q\right)$ and $r_{u}\left(x_{1}, a , Q\right)$ are the two roots of $r_{s}$ in the equation
\begin{eqnarray}
E\left( r_s ,a, Q\right)=E_{Z}\left(a , Q\right)  + x_{1}\left[E_{Y}\left(a , Q\right)- E_{Z}\left(a, Q \right)\right]. \label{rceqn1}
\end{eqnarray}
 The radii $r_{v}\left(x_{1}, a , Q\right)$ and $r_{u}\left(x_{1}, a, Q \right)$ obey $r_{v}\left(x_{1}, a, Q \right)> Z(a, Q)$ and $Y\left( a, Q \right)<r_{u}\left(x_{1}, a, Q \right)< Z\left( a, Q \right)$.

For the $\left( e, \mu \right)$ space, the corresponding formulae for the $\Delta$ region in Fig. \ref{emuQ2a2} are given by
\begin{subequations}
\begin{eqnarray}
e\left(x_{2} \right) =&& x_{2}, \label{ex} \\
\mu\left( y_{2}, r_{x}, a, Q\right)  =&&  y_{2}\cdot \mu_s\left( r_{x}, a, Q\right), \label{muy}
\end{eqnarray}
\end{subequations}
where the allowed range of $\mu$ is $0<\mu<\mu_s\left( r_{x}, a, Q\right)$ for a given $x_{2}$, and $\mu_s\left( r_{x}, a, Q\right)$ is the value of $\mu$ at the separatix [Eq. (25b) in Refs.~(\refcite{RMCQG2019,RM1arxiv2019})], and $r_{x}$ is a root of $r_{s}$ in the equation $x_{2}=e_s\left( r_{s}, a, Q\right)$ and $e_s\left( r_{s}, a, Q\right)$ is the eccentricity value at the separatix [Eq. (25a) in Refs.~(\refcite{RMCQG2019,RM1arxiv2019})]. The radius $r_x$ lies between $Y\left( a, Q \right)$ and $Z\left( a, Q \right)$ for a given $a$ and $Q$. Hence, for a fixed $a$ and $Q$, $ x_{2}$ and $y_{2} \in \left[ 0, 1 \right]$ which thereby defines the range of $\left( e, \mu \right)$.

\section{Summary and discussion}
\label{summary}
We presented the algebraic conditions for non-equatorial bound trajectories in the ($E$, $L$, $a$, $Q$) and ($e$, $\mu$, $a$, $Q$) spaces and showed how these conditions classify the bound orbits into various regions, $\Delta$, $\varsigma$, and $\Lambda$, in the ($E$, $L$) plane, which was previously discussed graphically in Ref.~\refcite{Hackmann2010}; see Fig. \ref{ELplane}. In this article, we have also shown these bound orbit regions in the ($e$, $\mu$) plane, Fig. \ref{emuQ2a2}, geometrically specified by their bound curves and vertices. For astrophysically relevant orbits, only the region $\Delta$ is applicable. We also provided a useful prescription to select the parameters ($E$, $L$) and ($e$, $\mu$) in the $\Delta$ region, which are canonical inputs to the trajectory solutions for studying their evolution in various applications like gravitational wave emission from extreme-mass ratio inspirals, relativistic precession around black holes, and the study of gyroscope precession as a test of general relativity.

We acknowledge the support from the SERB project CRG 2018/003415.

%\footnotesize

\end{document}